\documentclass[openacc]{rstransa}

\definecolor{egreen}{rgb}{0.0, 0.5, 0.0}
\definecolor{orange}{rgb}{0.96,0.60,0.06}

\begin{document}

\title{On the effect of oscillatory phenomena on Stokes inversion results}

\author{
P. H. Keys$^{1}$, O. Steiner$^{2,3}$ and G. Vigeesh$^{2}$}

\address{$^{1}$Astrophysics Research Centre, School of Mathematics and Physics, Queen's University Belfast, Belfast, BT7 1NN, U.K.\\
$^{2}$Leibniz-Institut f{\"{u}}r Sonnenphysik (KIS), Sch{\"{o}}neckstrasse 6, 79104 Freiburg, Germany\\
$^{3}$Istituto Ricerche Solari Locarno (IRSOL), Via Patocchi 57, 6605 Locarno-Monti, Switzerland}

\subject{astrophysics, wave motion}


\corres{Peter Keys\\
\email{p.keys@qub.ac.uk}}

\begin{abstract}
Stokes inversion codes are crucial in returning properties of the solar atmosphere, such as temperature and magnetic field strength. However, the success of such algorithms to return reliable values can be hindered by the presence of oscillatory phenomena within magnetic wave guides. Returning accurate parameters is crucial to both magnetohydrodynamics studies and solar physics in general. Here, we employ a simulation featuring propagating MHD waves within a flux tube with a known driver and atmospheric parameters. We invert the Stokes profiles for the 6301~{\AA} and 6302~{\AA} line pair emergent from the simulations using the well-known Stokes Inversions from Response functions (SIR) code to see if the atmospheric parameters can be returned for typical spatial resolutions at ground-based observatories. The inversions return synthetic spectra comparable to the original input spectra, even with asymmetries introduced in the spectra from wave propagation in the atmosphere. The output models from the inversions match closely to the simulations in temperature, line-of-sight magnetic field and line-of-sight velocity within typical formation heights of the inverted lines. Deviations from the simulations are seen away from these height regions. The inversion results are less accurate during passage of the waves within the line formation region. The original wave period could be recovered from the atmosphere output by the inversions, with empirical mode decomposition performing better than the wavelet approach in this task.
\end{abstract}


\maketitle

\section{Introduction}
Inversion algorithms are a useful tool in observational solar physics. By providing a means to infer various atmospheric parameters from Stokes profiles, one can begin to understand the evolution of physical properties in the solar atmosphere over time and/or with height depending on the observables. 

There are numerous inversion algorithms with various advantages and disadvantages associated with each based on the rationale for the actual code design. For example, Milne-Eddington inversions are fairly common (\cite{Landolfi1982, Landolfi1984, LitesSkumanich1990, OrozcoSuarez2010, Borrero2011, Bommier2016}; to name a few) and employ the Milne-Eddington assumption that the physical properties that govern line formation are constant with depth to solve the radiative transfer equation in the presence of a magnetic field \cite{Unno1956, Rachkovsky1962a, Rachkovsky1962b}. Milne-Eddington inversions tend to be quick to obtain a result, computationally speaking, and often give a good approximation of field strengths at a given height \cite{OrozcoSuarez2010, Bommier2007} though they do have difficulties dealing with asymmetric Stokes profiles due to the assumptions made by the code and the fact that real data may have gradients in the velocity and magnetic field components. 

More complex codes have been developed as well to fill the gap of specific needs for the community, such as depth varying Local Thermodynamic Equilibrium (LTE) based inversions utilising response functions of the observed signals \cite{RuizCobo1992, Frutiger2000}, codes that treat non-LTE conditions where the rate equations are not dominated by collisional terms \cite{SocasNavarro2000,SocasNavarro2015}, codes that account for both the Hanle and Zeeman effects \cite{Lagg2004, AsensioRamos2008}, codes that utilise archive synthetic spectra \cite{Beck2015}, non-LTE codes with analytical response functions \cite{Milic2018} and non-LTE codes that include partial redistribution effects \cite{delacruzRodriguez2019}. Each of these codes have different \textit{pro et contra}, with each offering a valuable and vital resource to the community for various use cases. More recently \cite{AsensioRamos2019,Carroll+Staude2001}, convolutional neural networks have been utilised with Stokes inversions to speed up the computational time taken to invert Stokes spectra. The results of this particular study would indicate that this is a useful approach to take and could be the direction that many inversion codes take in the future. 

Although different codes take different approaches to invert the data, in general they attempt to fit a model to an observed Stokes spectra for a given pixel. A common method to achieve this is to use a technique such as the Levenberg-Marquardt algorithm \cite{Press1989} to iteratively minimise the differences between post-radiative transfer profiles from a theoretical input model and the observed profiles through non-linear least-squares curve fitting. The model parameters (e.g, temperature, magnetic field, line-of-sight velocity) are then free to vary independently at several fixed points in the atmosphere (often referred to as nodes) until the differences are below some user specified value. The changes in-between these nodes are then assumed to behave as a polynomial, with a spline interpolation used to populate the model grid points between the nodes. Often the inversions are run on multiple cycles with the free parameters (i.e., the nodes) increased between cycles to improve the fits \cite{RuizCobo1992, SocasNavarro2011}. This nodal approach has been used in many inversion codes.


The utility of Stokes inversion algorithms has resulted in the outputs being employed in various studies (\cite{Balthasar2016, MartinezGonzalez2016, Henriques2017, Kuridze2018, Beck2019, DiazBaso2019, Keys2019, Kuckein2019, Keys2020}; to name some recent examples) and are a key technique in establishing and analysing height varying physical parameters in solar observations. In terms of wave phenomena the view is much less clear. One reason for this is due to the actual nature of the study, i.e., one is considering a periodically fluctuating atmosphere when a wave mode passes through. It is expected then that the oscillating atmosphere will perturb the parameters (e.g., temperature, B-field, density etc) to some degree due to the wave mode \cite{Solanki+Roberts1992,Fujimura2009, Moreels2013}, albeit these perturbations could be small. The selection of input model for the inversions then could be crucial, as the minimisation routine employed by the algorithm to fit the model atmosphere may overestimate or underestimate the contribution due to the wave phenomena on the Stokes parameter, which the routine is using to fit the model. The code may also attribute this to some other radiative transfer effect and miss the effect of the oscillation. 

This is not to say that wave studies utilising inversion algorithms are impossible to perform, they are seeing increasingly more usage in lower atmospheric wave studies \cite{Fujimura2009,delaCruzRodriguez2013,Stangalini2015, Henriques2017, Stangalini2018, Houston2018, Joshi2018}. However, extreme care needs to be taken in the analysis of the outputs of the inversion codes. This is particularly evident in the choice of free parameters in codes that employ the nodal approach to model fitting, where too many nodal points may create an erroneous artificial oscillatory phenomena as a result of the spline interpolation between nodes in the atmosphere. Likewise, there is a risk that a variation due to a real oscillatory motion may be misinterpreted as due to over fitting by the spline interpolation \cite{BellotRubio2000}. With all of this, there is an additional affect that needs to be considered, whereby the expansion of a flux tube with height may mean that within the line-of-sight of the observer, the observation may take in magnetic and non-magnetic (or rather less magnetic) regions within the one pixel, particularly in pixels close to the boundary of magnetic structures \cite{BellotRubio1996,Rezaei+al2007}. Some inversion codes have been adapted to try to account for such cases by allowing arbitrary stratifications of the various physical parameters \cite{Keller+al1990,BellotRubio2000}.

This can add complexity in the interpretation of sausage modes when the cross section of the oscillator may vary periodically and with height \cite{Grant2015, Freij2016, Keys2018}. The natural evolution of small-scale magnetic elements such as magnetic bright points (MBPs; \cite{Keys2019, Keys2020}) can further complicate the interpretation of wave signals in spectropolarimetric data as fluctuations can result from the natural evolution of these features as opposed to the propagation of waves. Another complication arises when the height of spectral-line formation oscillates in an otherwise static but height dependent magnetic field. This situation may lead to false detection of a magnetic oscillation \cite{Ruedi+Cally2003}. However, Vigeesh et al~\cite{Vigeesh2011} have shown the potential for using spectropolarimetric data in waves studies. In their study, the analysis of asymmetries in Stokes parameters within a simulated oscillating flux tube showed how observations of Stokes spectra could place a constraint on the different wave excitation mechanisms, particularly these smaller-scale network elements such as MBPs. One of the conclusions of the work, however, was that one needs adequate spatial and temporal resolution to resolve such asymmetries. In the era of the Inouye Solar Telescope (DKIST; \cite{DKIST}), this should become possible, however, here we put forward a study of whether current inversion algorithms can accurately detect and fit the Stokes parameters in an oscillating small-scale structure and whether they can accurately return the atmospheric parameters in the presence of an upwardly propagating wave.


\section{Data \& Methods}
\label{data}
The purpose of this work was to ascertain if commonly used inversion techniques could return accurate results, in the presence of a known oscillation. To do this we use a simple two-dimensional magnetohydrodynamic (MHD) model which was previously employed by Vigeesh et al.~\cite{Vigeesh2011}. The 2D initial atmosphere is in a Cartesian coordinate system and contains a magnetic flux sheet. The magnetic field configuration and pressure distribution is detailed by Vigeesh et al.~\cite{VigeeshHasan2009} and the model is constructed using the numerical methods outlined in Steiner et al.~\cite{SteinerPneuman1986}. Similar studies on these 2D models were employed previously \cite{Hasan2005, VigeeshHasan2009}. Full details of the simulations and model characteristics can be found in Vigeesh et al.~\cite{Vigeesh2011}.

The simulations, carried out with the code described in Steiner et al.~\cite{Steiner+al1998}, were generated for two field strengths (at $z = 0$) ranging between 1000\,G and 1600\,G on the axis of the sheet. The physical dimensions of the simulated domain were 1200~km in $x$ and 1300~km in $z$ with a pixel resolution of 5~km\,pixel$^{-1}$. The model advances by 2\,s in each frame for a total of 66 frames in the 1000\,G model and 41 frames in the 1600\,G model resulting in total durations of 132\,s and 82\,s for the 1000\,G and 1600\,G simulations, respectively. 

\begin{figure*}[h!]
\makebox[\linewidth]{
   \includegraphics[width=0.98\linewidth]{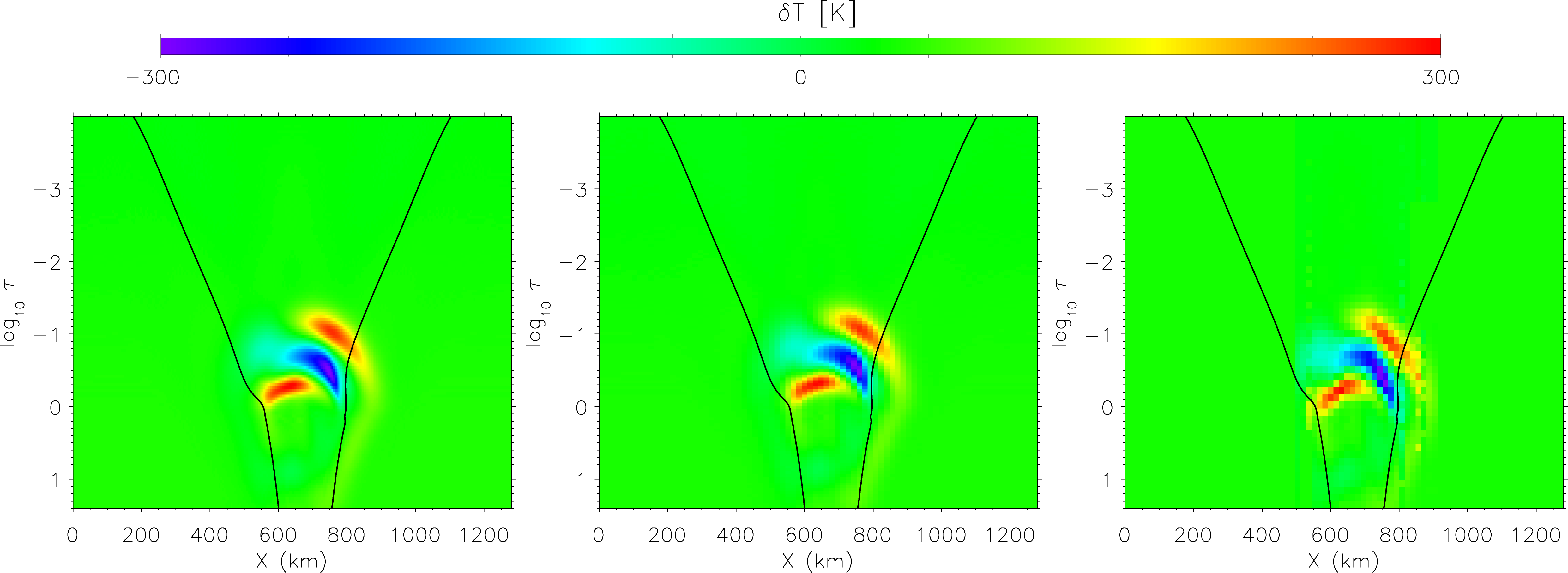}
	}
     \caption{Example time instant of the temperature perturbations for the 1000\,G simulation. The left panel shows the temperature perturbations in the original simulation while the middle panel shows the resultant temperture perturbations when the horizontal spatial resolution is degraded to a value expected for DKIST. The righthand panel shows the temperture perturbations derived from the inversion outputs for this snapshot. In all panels, the \textit{black} lines indicate the 100\,G contours of the flux sheet. The images show the effect of degraded spatial resolution on the retrieval of the temperature perturbations by the inversions and how the inversions were largely successful in reproducing the perturbations from the simulations.}
     \label{FigX}
\end{figure*}

An impulsive transverse excitation of the flux sheet at the lower boundary for both cases is simulated. A velocity amplitude of 5~km\,s$^{-1}$ and a period of 24\,s is specified at $z = 0$ following Equation (9) of Vigeesh et al.~\cite{Vigeesh2011}. Such transverse motions are not uncommon for small-scale magnetic elements \cite{Keys2011} and are due to the buffeting motion of granulation. The motion of MBPs is typically described by a random walk with occasional L{\'e}vy flights \cite{Cranmer2005, Keys2014, Jafarzadeh2017}, with this 5~km\,s$^{-1}$  burst corresponding to the L{\'e}vy-flight phase of an MBP's motion. This transverse driver generates magnetoacoustic waves within the flux sheet. The perturbation is 180$^\circ$ out-of-phase on the opposite sides of the sheet axis, leading to a quasi anti-symmetric wave pattern (see Figure~4 in Vigeesh et al.~\cite{Vigeesh2011}). At the $\beta = 1$ layer, these fast modes undergo mode transmission \cite{Cally2007} and change from fast to slow modes without changing their acoustic nature. When the wave vector is not exactly parallel to the magnetic field at the  $\beta = 1$ layer, partial conversion of the wave mode from fast acoustic to fast magnetic, reducing the energy in the acoustic mode \cite{VigeeshHasan2009}. An example of a temperature perturbation, in a similar style to Figure~4 of Vigeesh et al.~\cite{Vigeesh2011}, can be seen in the lefthand panel of Figure~\ref{FigX}.

Previously, Vigeesh et al.~\cite{Vigeesh2011} computed the emergent Stokes profiles from the top of the simulated box for both the moderate field (1000\,G) and strong field (1600\,G) simulations using the DIAMAG \cite{GrossmanDoerth1994} radiative transfer code for four Fe~{\sc{I}} lines. The authors find that, averaged over the whole simulated domain, the Stokes-$V$ profile does not show any significant variation with time, however, clear evidence of the wave propagation is observed in the emergent Stokes-$V$ profiles at higher spatially resolved lines-of-sight along the flux sheet. At higher resolution, the authors find clear asymmetries in the Stokes-$V$ profiles with opposite temporal behaviour on either side of the flux sheet central axis as a result of the quasi anti-symmetric wave pattern. The effects of refraction of the fast mode and following slow mode manifest as precursory and subsequent asymmetries and zero-crossing shifts of the Stokes-$V$ profiles as a function of time, respectively. Thus, Stokes-$V$ profiles can act as a diagnostic tool in wave studies.

Here we endeavour to take this idea further, by establishing how accurately common inversion algorithms can return key information from Stokes-$V$ profiles affected by wave phenomena. Given the conclusions of Vigeesh et al.~\cite{Vigeesh2011} with respect to spatially resolving these asymmetries, we decided to look at whether the new generation of solar telescopes such as the DKIST or the European Solar Telescope (EST; \cite{EST}) could spatially resolve the Stokes-$V$ asymmetries and whether currently available inversion codes could return the original atmosphere in the presence of these propagating fast modes. 

To that end, we took the simulations of Vigeesh et al.~\cite{Vigeesh2011} and we utilised the NICOLE \cite{SocasNavarro2011} code to synthesise the emergent Stokes profiles for both the moderate field and strong field simulations for the Fe~{\sc{i}} 6301~{\AA} and 6302~{\AA} line pair. We chose these lines as they are commonly used lines in spectropolarimteric observations of small-scale magnetic fields \cite{Utz2013, Stangalini2015, Jafarzadeh2015, Keys2019, Keys2020}. Also, it has been noted that inverting multiple lines in tandem, may improve the resulting outputs \cite{OrozcoSuarez2010, Riethmuller2019}, therefore, we wished to match as close as possible expected observation regimes for these facilities. The spectra were then degraded with a Gaussian blur in the $x$ axis to a typical diffraction limited resolution expected from DKIST. Here we use a spatial resolution in the $x$ axis of $\sim$30~km. Of course, this value is both instrument and wavelength dependent, however, we believe it is sufficient for our tests. An example of the degraded resolution simulation (the temperature perturbation in this case) can be seen in the middle panel of Figure~\ref{FigX}. We retain the temporal resolution in the first instance of 2\,s, however, again it should be noted that instrument choice for the spectropolarimteric observations may affect this value due to instrument scan times across typical lines. We performed the analysis on both the moderate and strong field simulations, as it has been seen previously \cite{Keys2019, Keys2020} that small-scale magnetic elements can exist in both regimes and can transition between the two frequently.

After producing the emergent Stokes profiles for both simulations, we employed SIR to run the inversions in LTE across the simulations for the degraded horizontal resolution. SIR was chosen for several reasons, 1) it has been used by the community for many years and has been shown to be stable, 2) the Fe~{\sc{i}} line pair we are inverting are photospheric lines, for which the assumption of LTE should be fairly valid in the inversion and 3) we tried to avoid using the same code to synthesise the Stokes spectra and invert them in case this added some form of bias on the subsequent inversions.\footnote{Note this is unlikely, but it seemed a prudent approach to take.} 

We treated the synthesised Stokes spectra as `observed' spectra and followed a similar approach that one would take in inverting spectropolarimteric data. We chose the HSRA quiet Sun model \cite{Gingerich1971} as our input model with an initial field strength of 1000\,G and a line-of-sight (LOS) velocity of 1~km\,s$^{-1}$. For many inversion codes the initial input model values for the magnetic field and LOS velocity component are randomised, however, with SIR these need to be specified. We chose these values based on typical values employed in the literature for these initial models, which is relevant as we are considering this as a hypothetical inversion of a small-scale magnetic field. We ran three inversion cycles, with increasing nodes and weighting in Stokes $V$ across the cycles, which are summarised in Table~\ref{Table1}. We only inverted the Stokes $I$ and $V$ profiles and ignored $Q$ and $U$. The reason for this decision is twofold: 1) this is a typical approach in quiet Sun observations, as the signal in $Q$ and $U$ is usually weak due to the polarimetric sensitivity of the instruments employed and 2) given this is a 2D model, we do not have the full vector magnetic field anyway, so it is likely that the results for such inversions would be spurious at least or difficult to interpret at best. An example of the success of the fits in Stokes $I$ and $V$ can be seen in Figure~\ref{Fig1} for both simulations. Also, an example of the temperature perturbation as derived from the inversion results can be seen in the righthand panel of Figure~\ref{FigX}.

\begin{table*}[h!]
\centering                                    
\caption{Nodes employed and weights used for Stokes parameters across inversion cycles.}         
\label{Table1}      
\begin{tabular}{l | c c c}         
\hline\hline
\textbf{Free Parameter} & \textbf{Cycle 1} & \textbf{Cycle 2} & \textbf{Cycle 3} \\
\hline                                   
Temperature & 2 & 5 & 7 \\
LOS Velocity & 1 & 3 & 5 \\
LOS B Field & 1 & 2 & 3 \\
\hline
\textbf{Weights} & & &   \\
\hline
Stokes $I$ & 1 & 1 & 1 \\
Stokes $Q$ & 0 & 0 & 0 \\
Stokes $U$ & 0 & 0 & 0  \\
Stokes $V$ & 5 & 10 & 10 \\
\hline                                            
\end{tabular}
\end{table*}

\section{Results \& Discussion}
\label{res}
The initial idea of this work was to establish if one could accurately retrieve the model atmosphere in an oscillating structure at the current best resolution available and, therefore, return more accurate information on the wave characteristics. The reason for this was to establish if such an approach is possible for real observations of small-scale magnetic fields and to determine what, if any, restrictions this places on the actual outputs for such studies. As such, after inverting the data, we initially decided to determine the success of the fits in Stokes $I$ and $V$ profiles from the inversions to the original input spectra. This is more important considering that the propagating wave results in asymmetries in the Stokes spectra \cite{Vigeesh2011}.

\subsection{Statistical metrics of Stokes profiles}
\label{res_stat_metrics}
To ascertain how well the Stokes spectra are modelled by our inversion regime, we take the coefficient of determination (R$^2$) value as a function of wavelength across the Stokes spectra independently for both the 1000\,G and 1600\,G simulations. The R$^2$ value is the square of the Pearson's product-moment correlation coefficient and represents the proportion of the variance within a dependent variable that can be predicted by the independent variables, thus, providing an estimate of how well the observed data are replicated by the output model. In this case, it predicts how well the synthetic Stokes spectra output by the inversion code are able to predict the `observed' profiles originally inverted. The closer the R$^2$ value is to one, the closer the model is able to predict the observed parameter, e.g., a value of 0.7 would suggest that the model explains 70\% of the variance in the observed data. There is some debate on the usefulness of R$^2$ \cite{Legates1999} as it can be sensitive to outliers in the data. However, as well as actually looking at the synthetic spectra with respect to the observed, the R$^2$ value offers at least some form of objective comparison of the accuracy of the inversions to reproduce the observed spectra and it offers a means to compare, to an extent, the success of the inversions between the data sets. The coefficient of determination can be explicitly calculated with the sum of the square of the residuals and the total sum of the squares with the following formula,
\begin{equation}
\mathrm{R}^2 = 1 - \frac{\sum_{i} (O_i - M_i)^2}{\sum_{i} (O_i - \overline{O})^2},
\end{equation}
where $O_i$ is the observed value, $M_i$ is the modelled value and $\overline{O}$ is the mean of the observed data.

\begin{figure*}[h!]
\makebox[\linewidth]{
   \includegraphics[width=0.93\linewidth]{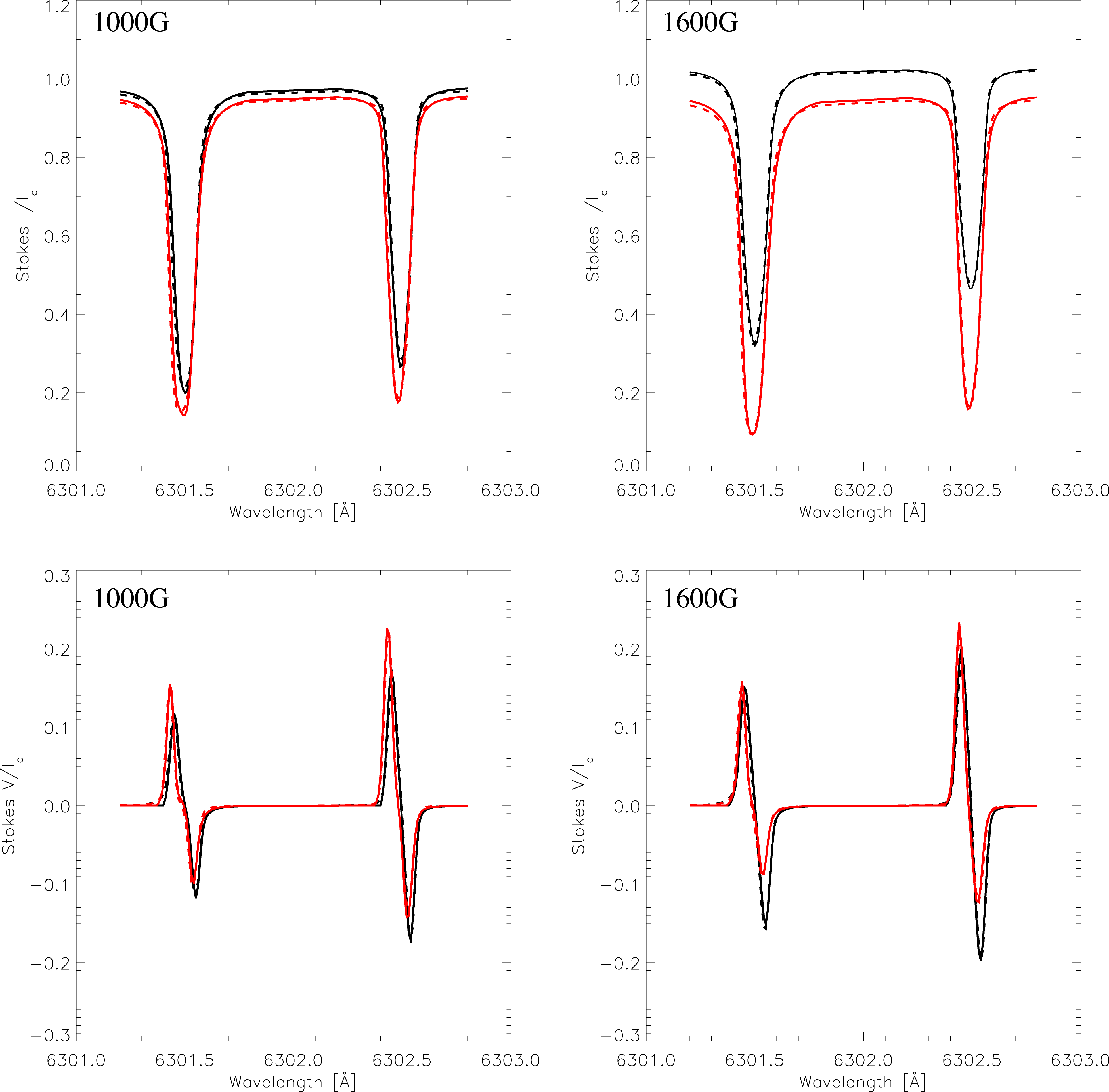}
	}
     \caption{The top row shows the Stokes-$I$ spectra for the 1000\,G (left) and 1600\,G (right) simulations, while the lower row shows the corresponding Stokes-$V$ spectra for the same simulations and for a line-of-sight parallel to but offset from the flux-sheet axis. The \textit{black} lines in each plot show the profile synthesised from the more quiescent phases of the simulations using NICOLE. These spectra were treated as `observed' spectra for the rest of the analysis. The \textit{black dashed} lines indicate the output Stokes spectra upon inversion of the `observed' spectra, showing the fits to the input data from our inversion set up. The \textit{red} lines in each case show an example of the Stokes $I$ and $V$ spectra in the presence of a propagating wave as synthesised by NICOLE, while the \textit{red dashed} line is the corresponding fit from our SIR inversions for these spectra after treating them as `observed' spectra. The Stokes-$V$ spectra in the perturbed case have a higher degree of asymmetry in the lobes, with a zero-point shift. The lines correspond to the photospheric Fe~{\sc{i}} 6301~{\AA} and 6302~{\AA} line pair.
}
     \label{Fig1}
\end{figure*}

The average R$^2$ value for the 1000\,G Stokes-$I$ spectra as a whole was 0.87$\pm$0.08 while the R$^2$ value for the Stokes-$V$ spectra was established as 0.71$\pm$0.11. With a value closer to 1 indicated a more precise fit of the observed spectra by the inverted data, it is clear for the 1000\,G case that the Stokes-$I$ spectra, on the whole, were more accurately returned by the inversion algorithm. For the 1600\,G case, the R$^2$ value for the Stokes-$I$ spectra as a whole was 0.91$\pm$0.09 while the R$^2$ value for the Stokes-$V$ spectra as a whole in this case was 0.82$\pm$0.15. For both the 1000\,G and 1600\,G simulations, the synthetic spectra appear to be able to predict the expected  observed spectra fairly well. This can also be seen in Figure~\ref{Fig1}, which shows sample Stokes $I$ and $V$ spectra (solids lines) and the corresponding synthetic spectra obtained through the inversions (dashed lines) for both the 1000\,G and 1600\,G simulations. 

In both the 1000\,G and 1600\,G cases, the Stokes-$I$ spectra appear to be more accurately replicated by the inversions. This is likely due to a few factors. Stokes $V$ is essentially the derivative of Stokes $I$, i.e., $V\propto {\rm d}I/{\rm d}\lambda$. When fitting both $I$ and $V$ together, it is plausible that it is easier to fit the function in the first place and harder to additionally fit its first derivative. One such factor is the typical relative scaling in values in the Stokes $I$ and $V$ spectra. That is, a weighting factor needs to be applied to the Stokes-$V$ spectra in the inversions to ensure that the Stokes-$V$ spectra are fit by the code more equally with respect to Stokes $I$. Given that this is a relatively large value, in our case 10 by the third cycle, small variations between profiles could make it more difficult for the inversion algorithm to reproduce the synthetic spectra accurately. Likewise, this is further complicated by asymmetries and zero point shifts in the profiles over time \cite{Vigeesh2011}. A combination of these factors will mean that the Stokes $V$ is probably less accurately reproduced than Stokes $I$ across all inverted profiles in the data, thus, resulting in a smaller R$^2$ value. 

It should be noted, that it is not advisable to make direct comparisons here between the R$^2$ values between the 1000\,G and 1600\,G simulations, to determine which case is more accurately reproduced by the inversion algorithm. The reason why this is not possible, is due to the fact that the two have different numbers of frames and, therefore, a different total number of spectra from which these parameters are obtained. Therefore, the 1600\,G R$^2$ value may be artificially higher in comparison to the 1000\,G  value purely for the fact that there are less measurements and potentially fewer outliers than the 1000\,G data. This would imply that the algorithm performs better in returning the synthetic spectra for the 1600\,G simulations, but that is not necessarily the case. A meaningful comparison cannot be made in this case. We can only specify that in both cases, the Stokes $I$ appears to  be more accurately reproduced by the inversion algorithm, however, we cannot say that the inversion algorithm reproduces the Stokes $I$ more accurately for the 1600\,G simulations in comparison to the 1000\,G simulations. With the same number of data points in the Stokes $I$ and $V$ spectra for a given simulation, we can make meaningful comparisons in comparing the outputs of the inversion algorithm in this scenario.

As stated previously, there is some debate on the usefulness of R$^2$ as a metric for determining model accuracy. There are several studies on this topic within the literature debating the issue \cite{Legates1999, Ritter2013, Li2016}. Adding to this debate is not the purpose of this study. In truth, many metrics used to determine model accuracy and goodness-of-fit have limitations and a degree of subjectivity from the assessor. To accurately assess the accuracy of a model output from SIR to the original simulation, a multi-objective approach is probably the best course of action \cite{Ritter2013}, employing various metrics and visually comparing the outputted synthetic Stokes profiles and model atmosphere to the originally inputted simulated profiles and the simulation used to generate them. As such, we also considered the dimensionless Nash and Sutcliffe \cite{NashSutcliffe1970} coefficient of efficiency (NSE), which is calculated using the root-mean-square-error (RMSE) and the standard deviation ($\sigma$) following,
\begin{equation}
NSE = 1 - \left( \frac{RMSE}{\sigma} \right)^2.
\end{equation}
The NSE has primarily been employed in the field of hydrology as an alternative to R$^2$ in evaluating model efficiency. However, given that it ostensibly determines the relative magnitude of the the residual variance to the measured data variance, it can also be employed here to give context to the goodness-of-fit of our inversions to the original observed data. The NSE can be sensitive to extreme values and can have values $-\infty \leq NSE \leq 1$, where a value of 1 represents a perfect fit and values below 0 indicate that the mean of the observed values is a better estimate than the model itself. Frequently a threshold of 0.65 and above is employed with NSE to indicate an acceptable model as this suggests that the model mean-squared-error represents 35\% of the observed variance.

For the 1000\,G simulations, the NSE value for the Stokes-$I$ profiles was 0.86$\pm$0.18, which is comparable to the results we obtained for the R$^2$ value in this case, suggesting good agreement between our model profiles output by the inversions and our observed profiles. Note, the values for R$^2$ and NSE for both simulations are summarised in Table~\ref{Table2}. For Stokes $V$, the NSE suggests a better fit in general at 0.85$\pm$0.12 in comparison to the R$^2$ value for the Stokes-$V$ profiles. However, both appear to suggest good agreement between our synthetic profiles output upon inversion with our original observations, with the synthetic profiles accounting for at least 70\% of the variance in our observed profiles. Similar results are obtained using the NSE with the 1600\,G simulations. The NSE for the Stokes-$I$ profiles was found to be 0.97$\pm$0.05 while the Stokes-$V$ profiles returned a NSE value of 0.91$\pm$0.08. It is difficult to make meaningful comparisons between the 1000\,G and 1600\,G simulations, however, the synthetic profiles extracted with SIR for the 1600\,G simulations appear to be more accurate to the input profiles. It is possible that this is due to a clearer definition of the profiles at higher field strengths which the code is able to recover more readily, but this is a somewhat speculative suggestion. Regardless, Stokes $I$ appears to fit better than Stokes $V$ for both simulation runs, however, the accuracy of the Stokes $V$ fits appears to be good enough in general to detect the asymmetries introduced by the upwardly propagating wave.

\subsection{Effect of wave propagation on Stokes fits}
\label{res_stokes_fit}

\begin{table*}[h!]
\centering                                    
\caption{Summary of Stokes fits statistics for all profiles, perturbed profiles and unperturbed profiles.}
\label{Table2}      
\begin{tabular}{c l | c c | c c}         
\hline\hline
 & & \multicolumn{2}{ c |}{\textbf{1000\,G}} &  \multicolumn{2}{c}{\textbf{1600\,G}} \\
\textbf{Statistic} & \textbf{Spectra} & Stokes-$I$ & Stokes-$V$ & Stokes-$I$ & Stokes-$V$\\
\hline                                 
 &  All spectra & 0.87$\pm$0.08 & 0.71$\pm$0.11 & 0.91$\pm$0.09 & 0.82$\pm$0.15 \\
R$^2$ & Perturbed spectra & 0.86$\pm$0.12 & 0.75$\pm$0.19 & 0.92$\pm$0.08 & 0.81$\pm$0.13\\
 & Unperturbed spectra  & 0.75$\pm$0.17 & 0.66$\pm$0.18 & 0.86$\pm$0.15 & 0.74$\pm$0.16\\
\hline
 & All spectra & 0.86$\pm$0.18 & 0.85$\pm$0.12 & 0.97$\pm$0.05 & 0.91$\pm$0.08 \\
NSE & Perturbed spectra & 0.88$\pm$0.15 & 0.82$\pm$0.13 & 0.97$\pm$0.04 & 0.89$\pm$0.10\\
 & Unperturbed spectra & 0.72$\pm$0.30 & 0.88$\pm$0.10 & 0.96$\pm$0.04 & 0.92$\pm$0.08\\
\hline    
\end{tabular}
\end{table*}

Now that we have ascertained that SIR was able to accurately reproduce the Stokes spectra for both simulations in general, we decided to look at how well the inversion could reproduce the Stokes spectra in the case of an emergent Stokes spectra that is affected by an upwardly propagating wave. Previously, the presence of a propagating wave was shown to induce an asymmetry and a zero point shift in the Stokes spectra \cite{Vigeesh2011}, so the inversion algorithm would need to accurately reproduce this in the synthetic spectra for the atmospheric model output by the algorithm to give confidence that the presence of the wave in the atmosphere is retained. To isolate the perturbed profiles from the unperturbed profiles, we found the time steps and locations that the upwardly propagating wave from the original simulations passed through the height of formation of the 6301~{\AA} and 6302~{\AA} line pair. We took this as an optical depth of $\log_{10} \tau = -0.5$ to $-1.5$, as studies suggest that this is roughly the region where the line forms \cite{MartinezGonzalez2006}. We took a range of optical depths to account for the fact that the optical depth is likely to vary in different regions of the tube and in the presence of the wave. Taking a range of optical depths ensures that we catch the propagation of the wave as it passes through that region of the atmosphere. Then we applied similar methods to profiles that were deemed as perturbed by the passage of the wave and those that were deemed unperturbed.

For the 1000\,G simulations, the R$^2$ value was 0.86$\pm$0.12 for the Stokes-$I$ spectra deemed to be perturbed and 0.75$\pm$0.17 for the unperturbed Stokes-$I$ spectra, while the NSE value was found to be 0.88$\pm$0.15 and 0.72$\pm$0.30 for the perturbed and unperturbed Stokes-$I$ profiles, respectively. The R$^2$ value for the corresponding Stokes-$V$ spectra for the 1000\,G simulation was 0.75$\pm$0.19 and 0.66$\pm$0.18 for the perturbed and unperturbed spectra, respectively. The NSE value for the perturbed Stokes-$V$ spectra was calculated as 0.82$\pm$0.13 while the unperturbed Stokes-$V$ spectra had an NSE value of 0.88$\pm$0.10. Note, Table~\ref{Table2} summarises the R$^2$ and NSE values for the perturbed and unperturbed spectra in both simulations.

Again, for both the Stokes $I$ and $V$ spectra, SIR appears to reproduce the Stokes $I$ more accurately than the Stokes-$V$ spectra. Interestingly, SIR appears to reproduce the perturbed Stokes $I$ and $V$ spectra more readily than the unperturbed cases with one metric (R$^2$), however, the opposite is shown by the NSE value though it should be noted that with the NSE value the goodness-of-fits for the perturbed and unperturbed cases appear to be very close. It is possible that SIR performs better with asymmetries within the Stokes spectra, however, the more likely reason for this is that there are fewer profiles that fall within the perturbed category given our selection criteria, than in the unperturbed category, which acts to artificially increase the R$^2$ value for the perturbed case. The NSE value may be less prone to this sample size issue, which may explain the discrepancy. Similar issues are noted for our inability to properly compare the success of the inversions in reproducing the spectra between the simulations so again it is difficult to make meaningful comparisons here on whether spectra in the perturbed case fit better or not through these means.

Similar trends are seen in the 1600\,G simulations. The R$^2$ value for the Stokes-$I$ profiles for perturbed profiles was 0.92$\pm$0.08 while the unperturbed Stokes-$I$ profiles had a R$^2$ value of 0.86$\pm$0.15. The corresponding NSE values for the perturbed and unperturbed Stokes-$I$ spectra was evaluated as 0.97$\pm$0.04 and 0.96$\pm$0.04, respectively. The Stokes-$V$ profiles for the 1600\,G simulation had a R$^2$ of 0.81$\pm$0.13 and 0.74$\pm$0.16 for the perturbed and unperturbed profiles, respectively. The NSE value for the perturbed Stokes-$V$ spectra and the unperturbed Stokes-$V$ spectra was evaluated as 0.89$\pm$0.10 and 0.92$\pm$0.08, respectively. Again we see the same issue here whereby the perturbed case appears to be closer to the observed profiles for both Stokes $I$ and $V$ in comparison to the unperturbed counterparts when considering R$^2$, whereas the NSE index indicates the opposite or at least that the fits are comparable for the perturbed and unperturbed spectra. The large value of R$^2$ for the perturbed Stokes-$I$ profiles reiterates the issue of smaller sample sizes in increasing R$^2$, as the 1600\,G simulations have fewer profiles and the perturbed profiles will be a smaller sample again. This does not detract from the fact that it points towards the fact that SIR has accurately reproduced the spectra, but it should be reiterated that it could account for the relatively larger R$^2$ values in the perturbed case here for the 1600\,G simulation. 

However, it should be noted that in both the perturbed cases and the unperturbed case for both the 1000\,G and 1600\,G simulations, the synthetic profiles match the observed profiles excellently, even accounting for the asymmetries and zero-point shifts in the profiles. This can be seen upon visual inspection of the synthetic and observed profiles. Figure~\ref{Fig1} shows examples of the synthetic and observed profiles for both the simulations and for the perturbed and unperturbed spectra in Stokes $I$ and $V$. The black lines indicate the unperturbed case while the red indicates and perturbed spectra. The dashed lines in each plot represent the synthetic profiles output by SIR upon inversion for each profile. Visual inspection of the profiles with the R$^2$ and NSE values suggests that the spectra are reproduced accurately by SIR even in the scenario were significant asymmetries are introduced by the effect of wave propagation on the spectra.

\subsection{Comparison of model parameters}
\label{res_model_comparison}
When we were confident that SIR is able to return accurate synthetic profiles for the Stokes $I$ and $V$ spectra for both the 1000\,G  and 1600\,G simulations, we turned our attention to the success of the output models from SIR with respect to the original simulations. A well fit profile from inversions does not necessarily mean that the output model atmosphere is an accurate representation of the actual solar atmosphere, as this can be affected by the choice of initial guess model and node values for the various parameters being fit and the fact that the inversion problem does not need to have a unique solution. We are in a unique position here, as we have the original simulations to compare with the output model from the inversions, however, we wished to treat this test as an observer would, which is why we matched common inversion regimes in inverting the data. In analysing the outputs we used similar techniques in comparing the inversion output to the original simulation in each time-step. We note that the inversion output does not have the same height scale as the original simulations and is limited to optical depths ranging from $\log_{10} \tau = 1$ to  $\log_{10} \tau = -4$, therefore, the success of the inversions is limited to this height range. This is acceptable as the line profiles that we synthesised from the simulations to invert form within this height region. Furthermore, we note that the spatial resolution with height of the simulations is significantly better than the model output from the inversions, however, we are still able to make meaningful comparisons between the atmosphere output by the inversions and the original simulation for various parameters.

In looking at the success of the model outputs from the inversions, we limited our analysis to the LOS velocity, the B$_{\mathrm{z}}$ component of the magnetic field and the temperature as these were the parameters in the inversion that were allowed to vary (i.e, we applied varying nodes per cycle in these parameters). Furthermore, these parameters would be of interest in wave studies, so it is prudent to ascertain the accuracy of these particular values given a known atmosphere. In general, the inversions were able to fit the atmosphere fairly well, albeit within a specific height region. The R$^2$ and NSE values for these three parameters help indicate that point on height related fits of the data.

For temperature, the R$^2$ value was calculated as 0.51$\pm$0.04 for the 1000\,G simulation and 0.49$\pm$0.07 for the 1600\,G simulations over the entire height scale of the inversions. The NSE value for the temperature in the whole inverted atmosphere was 0.55$\pm$0.10 and 0.53$\pm$0.08 for the 1000\,G and 1600\,G simulations, respectively. This would suggest that the output model from the inversions is slightly off the expected results from the original simulations. If we only consider a limited height range, i.e, from  $\log_{10} \tau$ = $-2.5$ to $\log_{10} \tau$ = $-0.5$, the results are improved slightly with an R$^2$ value of 0.56$\pm$0.07 and 0.55$\pm$0.03 for the 1000\,G and 1600\,G simulations, respectively. The NSE results suggest a better relation of 0.96$\pm$0.17 and 0.97$\pm$0.11 for the 1000\,G and 1600\,G simulations, respectively, in the same height range. 

\begin{figure*}[h!]
\makebox[\linewidth]{
   \includegraphics[width=0.98\linewidth]{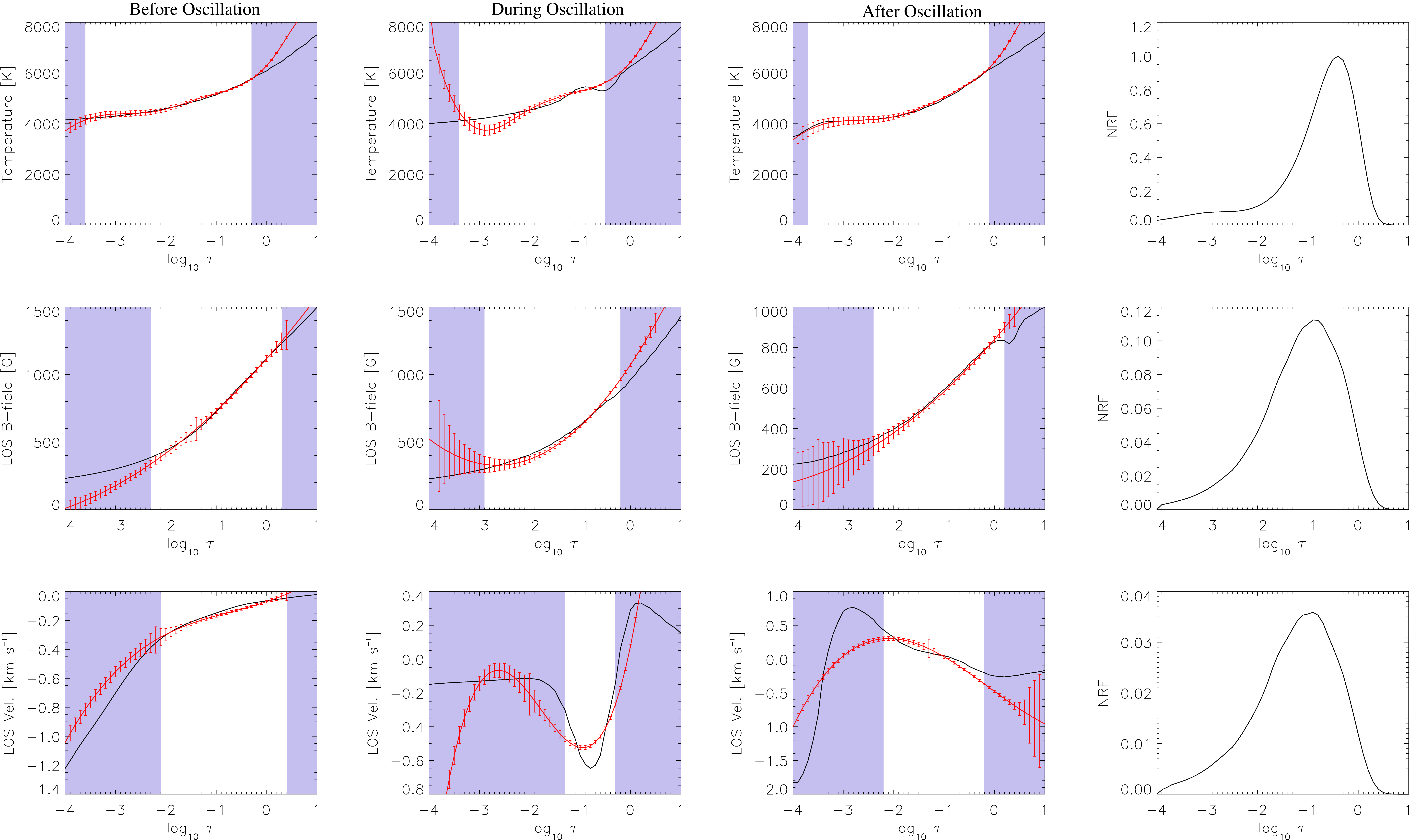}
	}
     \caption{Plots showing a comparison of the simulated atmosphere (\textit{black} line) with the atmosphere output by SIR for the corresponding location (\textit{red} line), just off the sheet axis of the 1000\,G model. The error bars from the SIR results are calculated with the response functions in an approach described by Socas-Navarro~\cite{SocasNavarro2011}. The top row shows the temperature, the middle row shows the LOS magnetic field and the bottom row shows the LOS velocity component. The first column shows the result before the wave has propagated through the region under analysis, the second column shows the atmosphere as the wave passes through the region and the third column shows the atmosphere some time after the wave has passed. The blue shading indicates regions where the R$^2$ value drops below 0.85, which suggests that the model less accurately represents the observations in these regions. The final column shows the normalised response functions for the three parameters (normalised to temperature) across the two lines, highlighting the optical depth of peak response for these parameters for comparison. The plots show that the inversions are fairly accurate representations of the real atmosphere within the height range of largest response of the spectral lines as seen in the plots in the final column. The plots also highlight the difficulty in accurately representing the atmosphere during the peak amplitude phase of an upwardly propagating wave, particularly in the case of the LOS velocity.
}
     \label{Fig2}
\end{figure*}

Similar results are obtained for both LOS B-field and LOS velocity, in that, when one considers a more limited height range, the models match closer to the original simulations. The R$^2$ value for the LOS B-field in this range was 0.81$\pm$0.03 and 0.84$\pm$0.02 for the 1000\,G and 1600\,G simulations, respectively. For the LOS velocity, the R$^2$ value in this height range was 0.82$\pm$0.02 and 0.79$\pm$0.03 for the 1000\,G and 1600\,G simulations, respectively. This is perhaps not all that surprising as one would expect the inversions to return a more accurate result within the heights that we would expect the highest contribution for the line being inverted. This is also reflected in what we see when we look at the inversion models in comparison to the original simulations, an example of which can be seen in Figure~\ref{Fig2}. We see that the inversion model approaches the original model in the heights typical of the 6301~{\AA} and 6302~{\AA} line pair (i.e., $\log_{10} \tau \approx -1$). Generally speaking in these regions, the output model matches the simulations and the general trends in these regions as well, within the errors associated with the inversion. This can be seen in Figure~\ref{Fig2} whereby the black line indicates the original simulation and the red line indicates the model atmosphere output by SIR. The error bars on the SIR plots are calculated using the response functions with the method described by Socas-Navarro \cite{SocasNavarro2011}. The errors bars give a visual indication of this reliability with height of the inversion results as well. Blue shading in Figure~\ref{Fig2} indicates regions that R$^2$ and NSE analysis suggest should be ignored, as the accuracy of the model to represent the observations is not sufficient.

Given the remit of this study, to determine if inversion algorithms could return accurate information in an oscillating atmosphere, we then decided to use a similar approach as used for the Stokes spectra and isolate the results for perturbed and unperturbed pixels. With this approach we again see that the three parameters are estimated with higher precision within the height of formation of the inverted line pair. As one might expect, the unperturbed cases across all three parameters return a closer fit to the original simulated atmosphere for both the 1000\,G and 1600\,G simulations. This can be seen in Figure~\ref{Fig2} which displays a case of the inversion results for the pre-perturbed atmosphere, during the peak amplitude phase of the perturbation and some time after the wave has passed the height that the lines typically form at. 

The inversion model returns a good fit, largely, in the perturbed case within the region of formation of the line, however, it tends to veer off from the expected values away from these regions quite noticeably in some instances. Fits were generally worse for the LOS velocity, with a relatively small acceptable height in the perturbed case. 

Another issue with the perturbed case, is that certain properties of the perturbed atmosphere (such as the sinusoidal like pattern in the LOS velocity in the second and third columns of Figure~\ref{Fig2}) are not fitted well, nor are they likely to ever be fitted well. For an observer with no \textit{a-priori} knowledge of the atmosphere, such a pattern would likely be ignored as over fitting by the inversion routine through the spline interpolation between node points. There is a danger that to return the values seen in the simulations, one may need to add more nodes, which increases the risk of overfitting the data and makes the interpretation harder for the observer. With that in mind, the inversion result does get an accurate value within the height of formation of the line for the LOS velocity in this case. Perhaps a multi-line and multi-height approach may give a clearer indication of these types of variations in the atmosphere in the case of propagating waves, particularly in the final panel of Figure~\ref{Fig2} when the velocity perturbation has made it to a higher region of the atmosphere. 

Another approach that could be considered to alleviate this issue is by improving the spatial resolution with height of the inversion. At present, it is likely that the height resolution of the inversion acts to smooth out some of these variations in the atmosphere due to the interpolation in the inversion. This is demonstrated in Figure~\ref{FigX}, whereby the left panel shows the temperature perturbation in the full resolution simulation, the middle panel shows the simulation degraded to the horizontal resolution of DKIST and the right panel shows the output from the inversions, which have lower spatial resolution in height with respect to the other two panels. It is clear to see in this image that the finer details of the perturbation are degraded in the inversion output. This is not unexpected, nor does it severely hinder the analysis of perturbations in this case. It may pose problems in the case of observations though, which will have a more complicated atmosphere to interpret.

Improving upon this is a difficult task due to the spatial scale of the perturbations (Figure~\ref{FigX}). If the perturbation is at a scale smaller than the mean free path of the photon, it is unlikely to be recovered in the inversion of a single line. In principle, the height resolution could be better than the photon mean free path by using a combination of various spectral lines with a range of formation heights, contribution functions and response functions. Clearly, this is not a simple task and requires careful consideration of observed spectral lines to achieve this, as well as considering the temporal resolution effects of a multi-line approach on the observations. This multi-line approach will be the subject of future work to ascertain how accurately such an approach can track an upwardly propagating perturbation in the atmosphere.


\subsection{The effect of wave propagation on model parameters}
\label{res_model_fit}
Now that we have established that the inversions return accurate parameters for both perturbed and unperturbed spectra within the region of formation of the spectral line in the atmosphere, we then turned our focus on whether the properties of the wave could be returned from the inversion result, specifically, could we return the wave period as determined from the temporal variation of the the physical state in the line formation region of the original model? 

The original simulations had a horizontal driver of 5~km\,s$^{-1}$ and a period of 24\,s, which, however, was driven for half a period (12\,s) only. Using the model atmosphere output by the inversions we then looked for the signatures of the wave over time by taking an average of the properties for each frame over the height region covering optical depths of $\log_{10} \tau = -0.5$ to $-1.5$. This is performed to smooth out any inconsistencies in the atmosphere from the inversions. With this approach we also ran some tests at a single optical depth ($\log_{10} \tau = -1$) to make sure that the results were comparable and that this averaging over 1~dex in height is a valid approach. The two approaches give comparable values in the end, retrieving periods within $\sim$2\,s for the corresponding signal. This is perhaps not surprising as the variation in the parameters in that height range will be somewhat minimal. We continued with the averaged approach as this is a typical technique used in inversion studies of properties with height.

We established the change in temperature, LOS B-field and LOS velocity within the regions specified in Vigeesh et al.~\cite{Vigeesh2011} (i.e., the regions marked either side of the flux sheet axis in Figure~4 of this previous study). Comparing the time varying signals to the corresponding signals from the original models, we were able to determine that the R$^2$ value was on average 0.87$\pm$0.02 for the three parameters between optical depths of $\log_{10} \tau = -0.5$ to $-1.5$. Therefore, in theory, we should be able to return the wave properties from the signals within this height range. 

To determine the properties of the wave we ran both wavelet analysis \cite{TorrenceCompo1998} and Empirical Mode Decomposition (EMD; \cite{Huang1998, Terradas2004, Huang2008}) on the change in temperature, LOS velocity and LOS B-field over the whole duration of the simulations. In general, we find that EMD performed better in returning the period of the oscillations, while the wavelet method had a tendency to underestimate it. For the 1000\,G simulation the wavelet returned an average period of 17.3$\pm$2.8\,s for the results from the inversions for temperature, LOS velocity and LOS B-field, while from the original model it returned an average period of 25.1$\pm$5.8\,s across the three parameters. For these simulations, for both the inverted results and the original model, the wavelet returns of the three parameters best results for the B-field. 

For the 1600\,G simulations, the wavelet method fails to return the wave period with any accuracy for both the inversion results and the original simulated atmosphere within the 99\% confidence level. It should be noted that the time series for the 1600\,G simulations is shorter in comparison to the 1000\,G simulations (with a duration of 82\,s in comparison to 132\,s), however, the wave frequency is still below the Nyquist frequency for the 1600\,G simulations, so it should still be capable of identifying the correct period in this case. Dropping the confidence level for these simulations to 95\% is still not sufficient to pick up the correct period. It should be noted that for both the 1000\,G and 1600\,G simulations, the wavelet also has a tendency to pick up a noise component with a period of $\sim 7 - 11$\,s within the confidence levels we have employed. It is of course possible that this could also come from the precursory fast wave.

By comparison, the EMD method returns more accurate result for both simulations in terms of the original wave period. For the 1000\,G simulations, EMD returns an average period of 24.9$\pm$2.0\,s for the original simulated atmosphere in temperature, LOS velocity and LOS B-field while for the same parameters obtained from the inversion outputs, EMD returns an average period of 23.1$\pm$1.3\,s. This is remarkably close to the original input driver period of 24\,s. Accurate periods for both temperature and LOS B-field were essentially returned with EMD, with the results from the LOS velocity slightly underestimated by EMD.

For the 1600\,G simulations, the results are not quite as good as the 1000\,G results for EMD, but are still considerably more accurate than the results returned for the 1600\,G simulations from the wavelet approach. With EMD, the average period for the original 1600\,G simulations was returned as 21.4$\pm$3.0\,s for temperature, LOS velocity and LOS B-field, while the corresponding result for the inversion outputs was 20.2$\pm$1.0\,s. Again, temperature and LOS B-field are returned more accurately for the 1600\,G simulations with EMD in comparison to the LOS velocity. 

\begin{figure}[h!]
\makebox[\linewidth]{
   \includegraphics[width=0.53\linewidth]{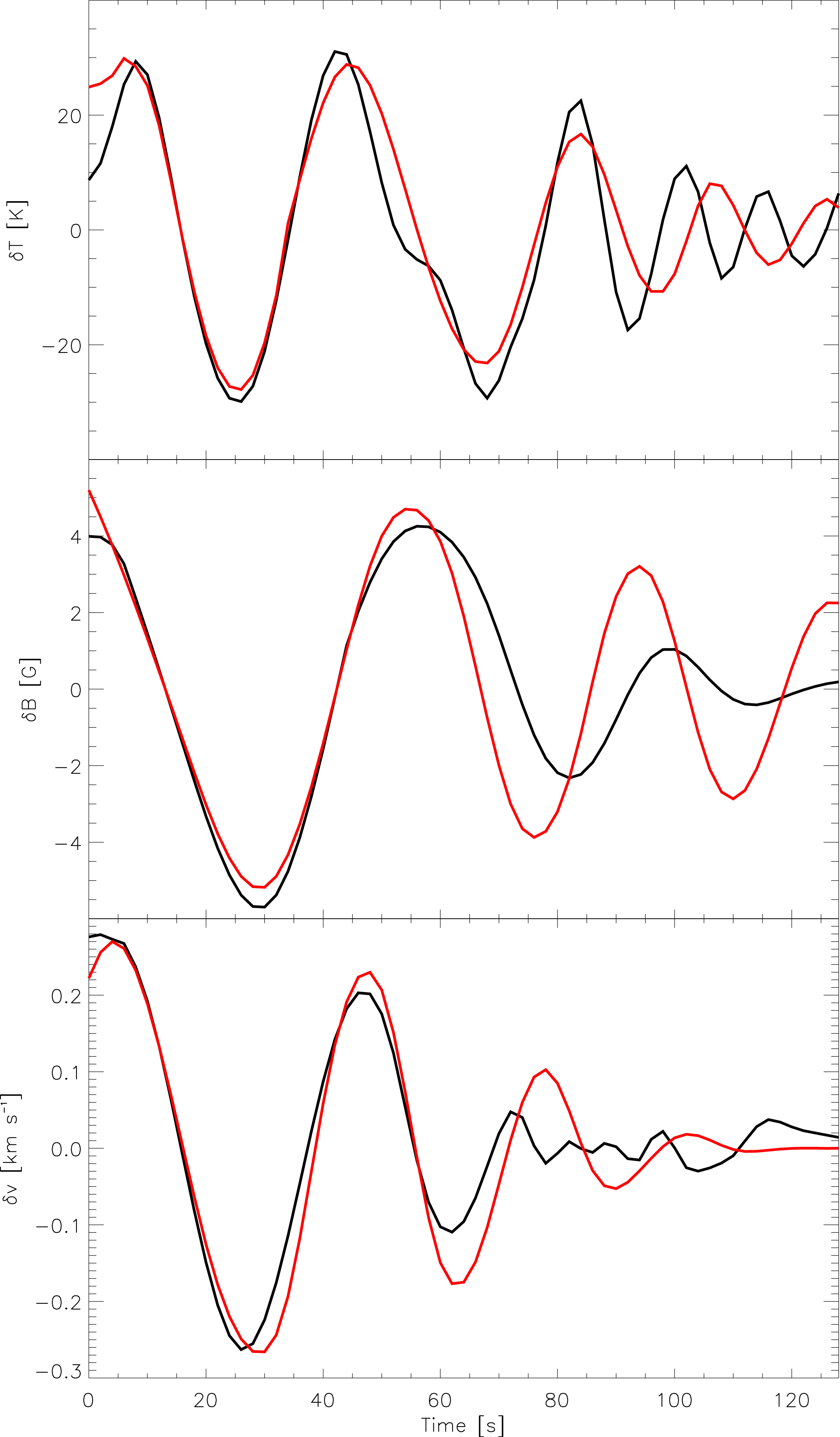}
	}
     \caption{Intrinsic Mode Functions (IMFs) obtained using the Empirical Mode Decomposition approach to find the properties of the original wave driver with a period of 24\,s. In each panel, the \textit{black} lines indicate the IMF for the original simulated atmosphere, while the \textit{red} line indicates the corresponding IMF for the atmosphere output by our inversions. All plots are for the 1000\,G simulations. The top plot shows the result from the analysis of the change in temperature, the middle plot shows the result from the analysis of the change in LOS B-field, while the bottom plot shows the results from the analysis of the change in LOS velocity. These plots correspond to a region just off the sheet axis and correspond to the region within between 410~km and 610~km in Figure~4 of Vigeesh et al.~\cite{Vigeesh2011}. The IMFs are able to pick out the 24\,s driver period towards the start of the simulations from the inverted atmosphere, which is reassuring that current nodal inversion techniques such as SIR can return the properties of an oscillating atmosphere sufficiently.
}
     \label{Fig3}
\end{figure}

An example of the Intrinsic Mode Functions (IMFs) from the three parameters in both simulations are shown in Figure~\ref{Fig3}, with the black curves showing the IMFs corresponding to the original simulated atmosphere and the red curves showing the corresponding IMFs for the inversion outputs in these parameters for the region between 410~km and 610~km in Vigeesh et al.~\cite{Vigeesh2011}, indicated by the blue solid lines in their Figure~4. It should be noted that EMD also picks out a noise component with a period of $\sim 7 - 11$\,s for both the 1000\,G and 1600\,G simulations. However, in the case of EMD, this noise component is returned as the zeroth IMF, which is commonly discarded as a noise element in the signal. With the wavelet approach, the noise element could be misinterpreted as a real oscillatory signal if due caution is not applied. It is not entirely clear why EMD performed better in returning the original wave period for the simulations for both the original simulations and the inversion outputs. It could be that the wavelet approach here is limited with low time-frequency resolution, which results in less accurate periods returned with the wavelet. From these results, EMD appears to be more accurate in returning the wave period for these small-scale magnetic elements and, therefore, may be useful in wave studies with MBPs. 

There are of course some limitations in this study that we should mention which we will address fully in future work. At present the inversions were run on the Stokes spectra that were generated for each frame, so that we have the full time resolution available from the simulations. However, with real observations (e.g., from a Fabry-P{\'e}rot instrument) there will be a lag in time across the line profile as the instrument scans the line. Typically the time for a line scan in reduced data for the lines employed here is about 10\,s for each line individually. Therefore, the asymmetries induced by the upwardly propagating wave may be varied across the spectra due to this time lag. Such an issue was investigated by Felipe~et~al.~\cite{Felipe2018,Felipe2019} with synthetic Stokes spectra of the chromospheric 8542~{\AA} line in the presence of umbral flashes. One study \cite{Felipe2018}, focussed on the effect of temporal resolution, while another \cite{Felipe2019} looked at the effect of spectral resolution. The authors find that fine spectral resolution ($\delta\lambda \approx 80$\,m{\AA}) is required for magnetic field inference, while line scanning within a time period of 16\,s can result in inversion results being unreliable in the early stages of an umbral flash. A great help to circumvent this problem could possibly consist in using an integral field unit (IFU) such as the DL-NIRSP instrument of DKIST, which makes scanning superfluous. 

Another issue with regards to real observations is the noise imparted on the spectra due to seeing conditions, so the spectra may have a noise component due to seeing variations during the scan time and also due to photon noise. It would be interesting to see how such noise components affect the ability of the inversion algorithms and, subsequently, the wavelet and EMD methods to return an accurate wave period for small-scale magnetic elements. The amplitude of the perturbations in magnetic field here are quite small (Figure~\ref{Fig3}) and increasing noise may act to mask out the perturbation signal. Here, we consider the LOS component of the magnetic field through the inversions of Stokes-$V$, which is proportional to the LOS magnetic field. For an accuracy of 1\,G, the polarimetric sensitivity will need to be greater than the noise level (i.e., $dV / V$), which will be approximately $1\times 10^{-3}$. For observed magnetic elements that are normally the focus of similar wave studies (e.g., MBPs) the Stokes-$V$ signal may be around 1\% of the continuum intensity. Therefore, the noise level in terms of continuum intensity ($d V / I_c$) for an accuracy of 1\,G would be around $1\times 10^{-5}$. This is a very challenging polarimetric accuracy for observations of similar perturbations, but in principle it is still achievable. Future work will look to add noise to the Stokes profiles prior to inversion to ascertain what noise level is tolerable to still retrieve the result seen in Figure~\ref{Fig3} with EMD for both the moderate and strong field simulations.

This study provided a preliminary analysis of the ability of current methods to return accurate model atmospheres in dynamic atmosphere with known input driver and atmospheric properties. We were able to return fairly accurate model parameters within the typical line formation heights for the lines studied at typical spatial resolutions expected with DKIST and EST. With the atmospheres output by the inversion algorithm, we were able to return the wave period of the original model more readily for the various parameters with the EMD approach. Given that we have seen in this preliminary study that these nodal inversion algorithms and EMD can return the atmosphere and wave period fairly accurately, future work will look at constraining their accuracy given typical observational limitations such as line scan times and seeing induced spectra noise as well as how multi-height inversions recreate the oscillating atmosphere. We expect that these considerations will hinder our ability to return an accurate wave period, however, we will analyse the extent of this in future work. Given the issues that we saw with passing perturbations in the LOS velocity, it is possible that a dedicated inversion algorithm may be needed for observational studies of waves in small-scale magnetic elements due to these conditions. The inversions here relied on static background models for the initial input, whereas with a dedicated wave inversion algorithm one could employ a more specific wave-driven model atmosphere which could potentially return a more accurate result in the presence of a perturbed atmosphere. 

Another possible solution for a dedicated wave inversion algorithm could be to use spectra from known simulated oscillations, in an approach similar to the CAlcium Inversion using a Spectral ARchive (CAISAR; \cite{Beck2015}) code, to return atmospheric parameters from asymmetrical spectra due to oscillations. In this approach, one could use simulated oscillations to produce spectra at various wavelengths and for various wave parameters to produce an archive of spectra. Then, similar to CAISAR, an algorithm finds the best match for the observed spectra that are being inverted to the synthetic archive spectra and the subsequent model atmosphere that produced the sythetic spectra. This could potentially return accurate atmospheric parameters in the presence of upwardly propagating perturbances in the atmosphere. Furthermore, CAISAR has been shown to be a fast inversion code (at less than 1\,s compute time per profile) so this method would also presumably return fast results to the user along with the underlining wave characteristics that produced the synthetic spectra for comparison to the observed spectra.

\section{Conclusion}
In this study, commonly used inversion algorithms were employed to look at the propagation of MHD waves from a known realistic horizontal driver in a two-dimensional magnetohydrodynamical simulation for both a moderate field (1000\,G) and strong field (1600\,G) flux sheet. The purpose of this was to ascertain whether inversion algorithms could fit the asymmetrical Stokes-$V$ profiles accurately and whether the output model from the inversion algorithm matched the model atmosphere used to generate the emergent Stokes spectra for the simulation. 

The emergent Stokes spectra were synthesised using the NICOLE inversion code for the Fe~{\sc{i}} 6301~{\AA} and 6302~{\AA} line pair. The spectra were then degraded to the same horizontal spatial resolution of the DKIST. These Stokes spectra were treated as `observed' profiles and inverted with the SIR inversion algorithm. The inversions were run with the SIR code with three cycles under the assumption of LTE. It was shown that the inversion process was entirely capable of fitting the Stokes-$V$ profiles, which were asymmetrical and had a zero-crossing shift through the propagating magnetohydrodynamic waves in the atmosphere, for a typical DKIST resolution image. 

The fitting of the Stokes spectra is crucial in returning accurate atmospheric values for a range of parameters (such as temperature, LOS velocity and B-field), therefore, this study then looked at the success of the inversion routine in returning these values, given that the original atmosphere was already known from the simulation. That is, how accurate was the returned atmospheric model when the Stokes spectra were inverted under the assumption of being a `real' observation. We find that the output model matches quite closely in both simulations over the height range of formation of the inverted lines, albeit, there is some deviation at the extreme ends of the model atmosphere, which is to be expected. This highlights the importance of multi-line spectropolarimetric observations of solar phenomena. Future work will look at simulated chromospheric spectra to decide how well this approach works in the case of wave studies.

For frames in which the MHD wave would pass through the typical formation heights of the line pair, the atmosphere is less accurately fit. This is unsurprising as the upwardly propagating wave perturbs the atmosphere over a height range smaller than the range over which the spectral lines are formed. We find that when the wave passes through the formation height of the lines, the LOS velocity is the least accurately returned parameter, with the inversion algorithm struggling to fit the perturbation. This highlights the issues faced in interpreting waves in real data. In terms of future wave studies, it would seem that a dedicated Stokes inversion code may be needed to accurately return physical parameters in the atmosphere, possibly by using a technique such as that used within the CAISAR code, with spectra for known oscillations generated by simulations similar to those presented here.

Finally, we considered whether wavelet and/or EMD could return the observed period in the model parameters of the original flux sheet, using the model atmospheres output by the inversions. We found that EMD performed better in this task, as the wavelet underestimated the period and returned a noise element as a possible real signal. However, we show that the original period can be obtained with these methods from the inversion results, which will be useful in the study of oscillations in small-scale magnetic elements.

\vskip6pt

\enlargethispage{20pt}


\dataccess{Reasonable requests for data and data products can be obtained by contacting the authors.}

\aucontribute{PHK ran and analysed the inversions. OS and GV generated the simulations. All authors interpreted the results.}

\competing{The authors declare that they have no competing interests.}

\funding{Research Council of Norway (project number 262622) and the Royal Society Theo Murphy Discussion Meeting (grant Hooke18b/SCTM)}

\ack{The authors wish to acknowledge scientific discussions with the Waves in the Lower Solar Atmosphere (WaLSA; www.WaLSA.team) team, which is supported by the Research Council of Norway (project number 262622) and The Royal Society through the award of funding to host the Theo Murphy Discussion Meeting `High resolution wave dynamics in the lower solar atmosphere' (grant Hooke18b/SCTM).}


\end{document}